\documentclass[prb,twocolumn,showpacs,amssymb,amsmath]{revtex4}

\usepackage{graphicx}
\usepackage{bm}
\def\be{\begin{equation}}
\def\ee{\end{equation}}
\def\bea{\begin{eqnarray}}
\def\eea{\end{eqnarray}}

\renewcommand{\vec}{\mathbf}
\usepackage{mathtools}
\usepackage{amsmath}
\usepackage{dsfont}
\usepackage{natbib}
\usepackage{color}

\newcommand{\Hamil}{\mathcal{H}}

\newcommand{\vS}{\vec{S}}

\begin{document}

\title{Functional-renormalization-group analysis of Dzyaloshinsky-Moriya and Heisenberg spin interactions on the kagome lattice}

\author{Max Hering$^1$}
\author{Johannes Reuther$^{1,2}$}
\affiliation{$^1$Dahlem Center for Complex Quantum Systems and Institut f\"ur Theoretische Physik,
Freie Universit\"at Berlin, Arnimallee 14, 14195 Berlin, Germany}
\affiliation{$^2$Helmholtz-Zentrum Berlin f\"{u}r Materialien und Energie, Hahn-Meitner-Platz 1, 14019 Berlin, Germany}
\date{\today}

\begin{abstract}
We investigate the effects of Dzyaloshinsky-Moriya (DM) interactions on the frustrated $J_1$-$J_2$ kagome-Heisenberg model using the pseudo-fermion functional-renormalization-group (PFFRG) technique. In order to treat the off-diagonal nature of DM interactions, we develop an extended PFFRG scheme. We benchmark this approach in parameter regimes that have previously been studied with other methods and find good agreement of the magnetic phase diagram. Particularly, finite DM interactions are found to stabilize all types of non-collinear magnetic orders of the $J_1$-$J_2$ Heisenberg model ($\vec{q}=0$, $\sqrt{3}\times\sqrt{3}$, and cuboc orders) and shrink the extents of magnetically disordered phases. We discuss our results in the light of the mineral {\it herbertsmithite} which has been experimentally predicted to host a quantum spin liquid at low temperatures. Our PFFRG data indicates that this material lies in close proximity to a quantum critical point. In parts of the experimentally relevant parameter regime for {\it herbertsmithite}, the spin-correlation profile is found to be in good qualitative agreement with recent inelastic-neutron-scattering data.
\end{abstract}

\maketitle

\section{Introduction}
According to a more traditional understanding of solid-state physics, the effects of spin-orbit coupling (SOC) are small relativistic corrections that can be neglected in most materials. However, the recent synthesis of a growing number of materials where SOC is a non-negligible order-one effect~\cite{Kim2008,Plumb2014,Plumb2015,Gegenwart2015} has substantially changed this perspective. In magnetic systems, SOC generally leads to anisotropic spin interactions that may induce novel types of quantum phases and quasiparticles. A famous example is the analytically solvable Kitaev model on the honeycomb lattice with its characteristic bond-dependent Ising interactions, giving rise to a spin liquid phase and emergent Majorana excitations~\cite{Kitaev2006}. Possible candidate materials~\cite{Jackeli2009,Singh2010,Chaloupka2010} to realize such physics in nature exhibit heavy magnetic ions which increase the magnitude of SOC. Another type of magnetic anisotropy induced by SOC is the off-diagonal and antisymmetric DM interaction \cite{Dzyaloshinsky1958,Moriya1960} which does not primarily depend on the atomic number $Z$ but crucially relies on the lattice geometry. It appears whenever the center of a bond connecting two magnetic ions is not an inversion center of the underlying lattice. In contrast to the anisotropic Ising interactions of the Kitaev model, DM couplings usually induce magnetic orders of non-collinear type and may stabilize exotic spin arrangements such as spiral orders or skyrmions \cite{Muhlbauer2009,Nagaosa2013}.

One of the simplest two-dimensional lattices where DM exchange is a symmetry-allowed interaction even on nearest-neighbor bonds is the kagome lattice, see Fig.\ \ref{fig:DMkagome}(a). Built of a network of corner-sharing triangles, it is at the same time a paradigmatic example for a strongly frustrated lattice. There is indeed a wealth of evidence from different numerical methods that the antiferromagnetic nearest-neighbor spin-1/2 Heisenberg model on the kagome lattice features a magnetically disordered ground state which might even realize a quantum spin liquid~\cite{Ran2007,Jiang2008,Yan2011,Lu2011,Iqbal2011,Li2012,Messio2012,Iqbal2013,Suttner2014,Rousochatzakis2014,Kolley2015,Mei2016}. Given its tendency to induce magnetic order, the DM interaction is, hence, an important perturbation of the kagome lattice that could alter the ground state significantly. One of the prime questions is whether and at which strength the DM exchange can destroy the presumed spin-liquid phase. Numerical studies such as exact diagonalization indicate that the non-magnetic phase survives up to a ratio of the nearest-neighbor DM and Heisenberg interactions of $D/J_1=0.1$, giving way to a magnetically ordered ${\bf q}=0$ state above this value~\cite{Cepas2008,Rousochatzakis2009,Seman2015}.

Apart from its theoretical importance as a generic frustrated spin system, there is also a growing number of material realizations for the kagome lattice. Currently, the cleanest implementation of an antiferromagnetic nearest-neighbor Heisenberg model on the kagome lattice is the mineral \textit{herbertsmithite} ($\mathrm{ZnCu_3(OH)_6 Cl_2}$) \cite{Shores2005,Han2012,Han2016}, which consists of weakly coupled kagome planes of spin-$1/2$ copper ions~\cite{Jeschke2013}. Most importantly, the absence of any long-range magnetic order down to $50 \, mK$~\cite{Shores2005, Mendels2007, Helton2007} in conjunction with a very broad spinon-like excitation spectrum~\cite{Vries2009,Han2012,Han2016} renders {\it herbertsmithite} one of the most promising spin-liquid candidates synthesized so far. Concerning the size of the DM interaction, ESR measurements imply a relative strength of $D/J_1=0.08,...,0.1$~\cite{Zorko2008} which, interestingly, puts this mineral exactly into the parameter regime where theory predicts the onset of magnetic order. Therefore, the DM coupling may drive {\it herbertsmithite} very close to a quantum critical point, raising questions about the precise location of the phase boundaries in the experimentally relevant parameter range.

In this article, we study the effects of DM interactions on the spin-$1/2$ kagome-Heisenberg model using the PFFRG method which has proven to accurately describe magnetic and non-magnetic phases of frustrated quantum spin systems~\cite{Reuther2010,Singh2012,Suttner2014,Iqbal2016,Buessen2016}. To this end, we extend the existing PFFRG technique to treat systems with finite DM couplings. As shown below, the off-diagonal nature of the DM interaction generates additional vertex functions with reduced symmetries which complicates a PFFRG analysis enormously as compared to diagonal exchange interactions. Despite the increased computational effort for numerically evaluating the renormalization-group equations, we reach sufficiently large system sizes and frequency resolution to appropriately describe the combined effects of Heisenberg and DM interactions. Particularly, as a first test of its applicability, we find that the critical ratio of $D/J_1\simeq0.1$ for the onset magnetic order is well reproduced, indicating that the accuracy of the PFFRG is retained when finite DM interactions are added. To put the $J_1$-$D$ model on the kagome lattice into a broader context, we also study the full $J_1$-$J_2$-$D$ model, where $J_2$ is the second-neighbor Heisenberg interaction, and $J_1$ and $J_2$ can both be ferromagnetic or antiferromagnetic. The motivation for this type of extended model comes from ab initio calculations for \textit{herbertsmithite} which predict a small antiferromagnetic $J_2$ coupling given by $J_2/J_1\simeq 0.019$~\cite{Jeschke2013}. Within PFFRG, we find that the DM interaction increases the size of all non-collinearly ordered phases of the original $J_1$-$J_2$-Heisenberg model (i.e., $\sqrt{3}\times\sqrt{3}$, $\vec{q}=0$, and cuboc order) but leaves the ferromagnetic phase unaffected. In parameter regimes which are experimentally relevant for \textit{herbertsmithite}, we qualitatively reproduce the correlation profile of recent neutron-scattering experiments\cite{Han2012,Han2016}. However, we also find small but non-negligible indications of magnetic order in these regimes which might imply that additional sources of frustration are needed to fully capture the microscopic situation in this material.

The paper is organized as follows: In Sec. \ref{sec:TheModel}, we introduce the microscopic model and fix our convention for the DM interaction. Sec. \ref{sec:FRG} then outlines the essentials of the PFFRG approach~\cite{Reuther2010,Metzner2012}, where Sec.~\ref{sec:FRGA} first gives a brief introduction into the general PFFRG framework for Heisenberg systems while Sec.~\ref{sec:FRGB} discusses the modifications for finite DM interactions. Thereafter, we investigate the $J_1$-$D$ model on the kagome lattice in Sec. \ref{sec:NearestNeighbor}. To gain a better understanding of the formation of ${\bf q}=0$ order in this model, we first solve the flow equations analytically in a limit where the PFFRG reduces to the classical random phase approximation (RPA). We further discuss the full $J_1$-$J_2$-$D$ model in Sec. \ref{sec:NextNearestNeighborA} and show how the DM interaction changes the magnetic phase diagram of the $J_1$-$J_2$-Heisenberg model. Parameter regimes relevant for {\it herbertsmithite} are investigated in Sec. \ref{sec:NextNearestNeighborB} and the results are compared to neutron-scattering data. Finally, Sec.~\ref{sec:Conclusion} contains a summary and conclusion of the entire work.

\section{Microscopic model}
\label{sec:TheModel}
\begin{figure}[t]
\includegraphics[width=3.1in]{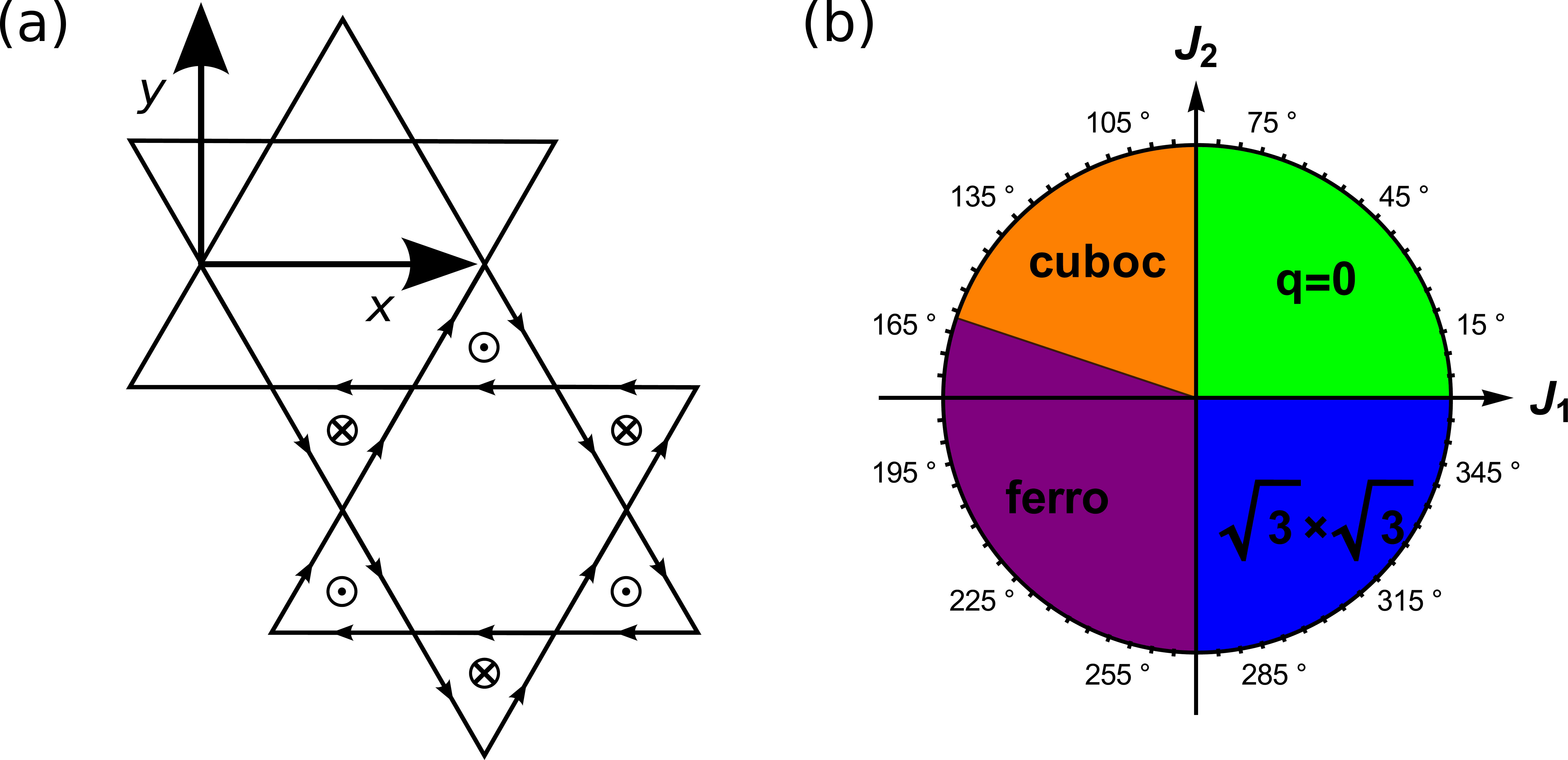}
\caption{(Color online) (a) Illustration of the DM interaction on the kagome lattice: Arrows on the nearest-neighbor bonds indicate the bond orientation of the DM term $\sim\vec{D}_{ij}\cdot\left(\vec{S}_i \times \vec{S}_j \right)$. Each arrow starts at site $i$ and ends at site $j$. For up-pointing (down-pointing) triangles the DM vector $\vec{D}_{ij}$ is oriented parallel (antiparallel) to the $z$ axis. (b) Classical phase diagram of the $J_1$-$J_2$-Heisenberg model featuring $\vec{q}=0$ N\'{e}el order, cuboc order, ferromagnetic order, and $\sqrt{3}\times\sqrt{3}$ N\'{e}el order: The transition between the cuboc and the ferromagnetic phase occurs at $J_2=-J_1/3$ with $J_1<0$ while all other phase transitions coincide with the $J_1$ or $J_2$ axis, respectively\cite{Domenge2005}.   \label{fig:DMkagome}}
\end{figure}

The Hamiltonian of the $J_1$-$J_2$-$D$ model studied in this article is given by
\begin{equation}
 \label{eq:NNNHamiltonian}
 \Hamil= J_1\sum_{\langle i,j\rangle} \vS_i \cdot \vS_j + J_2\sum_{\langle\langle i,j\rangle\rangle} \vS_i \cdot \vS_j + \sum_{\langle i,j\rangle} \vec{D}_{ij} \cdot \left ( \vS_i \times \vS_j \right),
\end{equation}
where $\langle i,j\rangle$ are nearest-neighbor pairs and $\langle \langle i,j\rangle\rangle$ denotes second-neighbor pairs of sites. According to Moriya's rules~\cite{Moriya1960}, the DM term $\sim\vec{D}_{ij} \cdot\left ( \vS_i \times \vS_j \right)$ is a symmetry-allowed coupling on nearest-neighbor kagome bonds since the bond center is not an inversion center of the lattice. Furthermore, the vector $\vec{D}_{ij}$ must be aligned perpendicular to the system's mirror plane which, in our case, is the kagome plane itself. Due to $\vec{D}_{ij}=-\vec{D}_{ji}$, the DM interaction defines an orientation of the bonds which we choose as shown in Fig.~\ref{fig:DMkagome}(a). With this convention, the point-group symmetries of the kagome lattice fix the directions of the DM vectors such that $\vec{D}_{ij}=\pm D \vec{e}_z$ is oriented parallel (antiparallel) to the $z$ axis on up-pointing (down-pointing) triangles [see Fig.\ \ref{fig:DMkagome}(a)], or vice versa. Up to small tilts of $\vec{D}_{ij}$ into the $x$-$y$ plane, this is also the relevant configuration for \textit{herbertsmithite}. It is worth noting that the presence of the DM term breaks the $SU(2)$ spin-rotation symmetry down to $U(1)$ rotations around the $z$ axis which, in combination with the off-diagonal nature of the DM coupling, requires significant adjustments of the PFFRG procedure.

The nearest-neighbor $J_1$-$D$ model with antiferromagnetic $J_1$ has previously been investigated by C\'{e}pas {\it et al.} \cite{Cepas2008} employing exact diagonalization. They find that the magnetically disordered phase is sustained for small DM couplings up to a critical ratio of $D/J_1\simeq0.1$ while the system is driven into a $\vec{q}=0$ N\'{e}el ordered phase for stronger DM interactions. Concerning the sign of the DM coupling, it can be shown that models with positive and negative $D$ can be mapped onto each other (e.g., by performing a global $\pi$-rotation of all spins in the $x$-$z$ plane). For the $\vec{q}=0$ state, this means that switching the sign of $D$ reverses the chirality of the spin orientations but does not change the spin-spin correlations. We therefore restrict ourselves to the case $D\geq0$ in the following.

The pure Heisenberg $J_1$-$J_2$ model on the kagome lattice has previously also been studied by various methods including PFFRG~\cite{Suttner2014,Iqbal2015,Kolley2015,Buessen2016}. Classically, this model supports four magnetically ordered phases referred to as $\vec{q}=0$ N\'{e}el order, cuboc order, ferromagnetic order, and $\sqrt{3}\times\sqrt{3}$ N\'{e}el order (for real-space illustrations of these types of orders we refer the reader to Ref.\ \citenum{Messio2011}). The classical phase diagram is shown in Fig.\ \ref{fig:DMkagome}(b) and the corresponding positions of the dominant susceptibility peaks in $\vec{k}$ space are depicted in Fig.\ \ref{fig:NNSus}(a). As a results of quantum effects, two extended magnetically disordered phases are found to emerge around $(J_1,J_2)=(1,0)$ and  $(J_1,J_2)=(0,1)$~\cite{Suttner2014,Gong2015,Buessen2016}. The $J_1$-$J_2$-$D$ model for antiferromagnetic $J_1$ and $J_2$ interactions has been studied by Seman {\it et al.} \cite{Seman2015} within exact diagonalization, predicting gapped and gapless spin liquid regimes in the quantum-disordered phase of the model. In this work, we complete the analysis of the $J_1$-$J_2$-$D$ model by also allowing for ferromagnetic Heisenberg couplings.

\section{Functional renormalization group for spin systems}
\label{sec:FRG}
The PFFRG method has proven to be remarkably accurate in describing the interplay between magnetically ordered and disordered phases in frustrated quantum-spin models. So far, this approach has mostly been applied to $SU(2)$ spin-rotation-invariant Heisenberg models on various 2D and 3D lattices~\cite{Reuther2014,Suttner2014,Balz2016,Buessen2016}. Extensions for anisotropic but {\it diagonal} exchange couplings are relatively straightforward and have been employed to study Kitaev interactions on the honeycomb lattice~\cite{Singh2012,Reuther2014a} and to resolve spin-nematic types of long-range order~\cite{Iqbal2016a}. In contrast, the implementation of anisotropic and {\it off-diagonal} DM interactions, as presented below, is found to be more involved and has so far not been attempted within PFFRG. Before we explain all necessary modifications of the approach in Sec. \ref{sec:FRGB}, we first briefly review the general PFFRG setup in the case of Heisenberg interactions. 
 \begin{figure*}[t]
\includegraphics[width=6.0in]{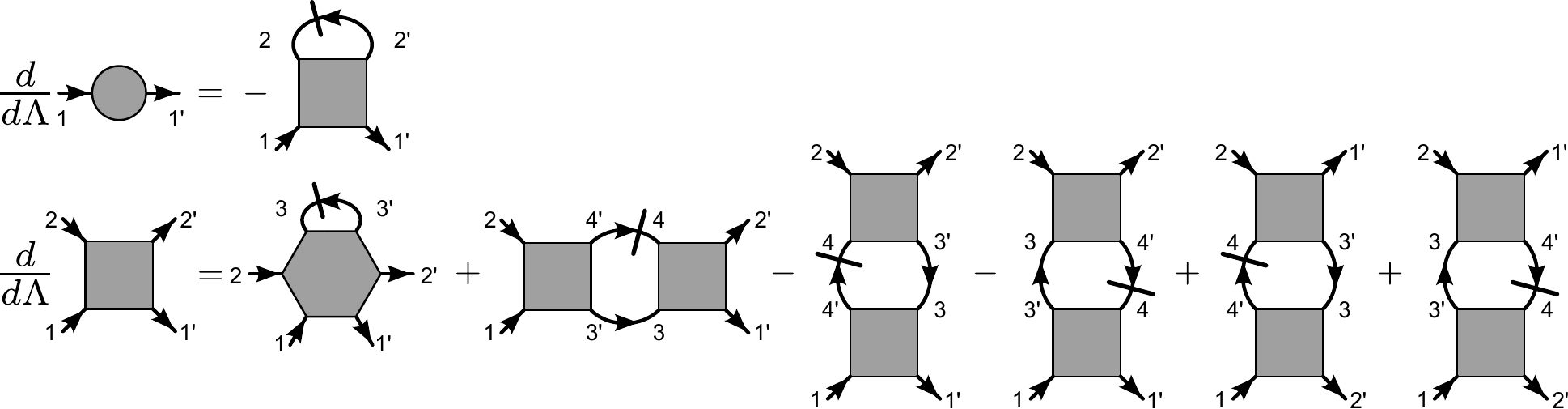}
\caption{Diagrammatic representation of the FRG equations in Eqs.\ \eqref{eq:FirstFlow} and \eqref{eq:SecondFlow}: The $m$-particle vertices are illustrated as gray shaded disks or polygons. Arrows with (without) a slash denote the single-scale propagator $S^\Lambda$ (fully dressed propagator $G^{\Lambda}$). The flow of the self energy $\Sigma^\Lambda$ couples to itself via $S^{\Lambda}$ and to the two-particle vertex $\Gamma^\Lambda$ (first line). In a similar fashion, the flow of the two-particle vertex $\Gamma^\Lambda$ couples to all $m$-particle vertices with $m \leq 3$ (second line). This scheme repeats for higher vertices up to infinite order (not shown). \label{fig:FRGEquations}}
\end{figure*}

\subsection{General PFFRG approach for Heisenberg systems}\label{sec:FRGA}

The starting point of the PFFRG procedure amounts to rewriting the spin operators from Eq. \eqref{eq:NNNHamiltonian} in terms of pseudo fermions to be able to employ standard fermionic diagram techniques,
\begin{equation}
\label{eq:AuxiliaryFermions}
 S^{\mu}_i=\frac{1}{2}\sum\limits_{\alpha,\beta} f^{\dagger}_{i,\alpha}\sigma^{\mu}_{\alpha\beta}f_{i,\beta}.
\end{equation}
Here, $\sigma^{\mu}_{\alpha\beta}$ with $\mu \in \{x,y,z\}$ are the Pauli matrices and $f_{i,\alpha}$ ($f^{\dagger}_{i,\alpha}$) denote annihilation (creation) operators of a spin-$\alpha$  fermion ($\alpha=\uparrow,\downarrow$) at lattice site $i$. While the physical spin states $\uparrow$ and $\downarrow$ are realized in the subspace with one fermion per lattice site ($Q_i\equiv f^{\dagger}_{i,\uparrow}f_{i,\uparrow}+f^{\dagger}_{i,\downarrow}f_{i,\downarrow}=1$), the fermionic representation also leads to spurious doubly ($Q_i=2$) or non-occupied ($Q_i=0$) states which do not carry a spin. A simple approximative scheme to fulfill the pseudo-fermion constraint $Q_i=1$ is to enforce its thermodynamic average $\langle Q_i\rangle=1$ which amounts to introducing a chemical potential $\mu_{i,\alpha}$ acting on the fermionic system. Due to the particle-hole symmetry of Eq.~\eqref{eq:AuxiliaryFermions}, this chemical potential vanishes identically throughout the lattice. While the average constraint $\langle Q_i\rangle=1$ in principle still allows for local particle-number fluctuations, it can be shown that states with $Q_i=0,2$ are associated with a finite excitation energy on the order of the exchange couplings~\cite{Baez2016} such that unphysical occupations are completely suppressed at $T=0$.

The basic building block of a diagrammatic theory for Eq.~\eqref{eq:NNNHamiltonian} is the free fermionic propagator $G_0$ given by
\begin{equation}\label{eq:BareGF}
G_0=\frac{1}{i\omega_n},
\end{equation}
where $\omega_n$ denotes the Matsubara frequency. It is worth emphasizing that due to the absence of any quadratic term in the fermionic version of Eq.~\eqref{eq:NNNHamiltonian}, the free propagator is local in real space and diagonal in spin space (note that the locality is also retained in all orders of diagrammatic expansions).

Within PFFRG, this propagator is regularized by a step function which suppresses the fermionic propagation in a frequency interval between $\omega_n=-\Lambda$ and $\omega_n=+\Lambda$,
\begin{equation}
 \label{eq:LambdaGF}
 G_0^{\Lambda}=\theta\left ( \left | \omega_n  \right | - \Lambda \right) G_0,
\end{equation}
where $\Lambda$ is the so-called RG scale. With this modification, the generating functional for the one-particle irreducible $m$-particle vertices becomes $\Lambda$ dependent. According to the standard FRG framework~\cite{Metzner2012,Enss2005, Hedden2004, Salmhofer2001, Wetterich1993}, the scale derivative of the generating functional yields an infinite hierarchy of integro-differential equations where the $\Lambda$ flow of each $m$-particle vertex couples to all $n$-particle vertices with $n\leq m+1$. The first two FRG flow equations for the self energy $\Sigma$ and the two-particle vertex $\Gamma$ read
\begin{align}\label{eq:FirstFlow}
 \frac{d}{d\Lambda}\Sigma^{\Lambda}\left(1';1\right)&=-T\sum\limits_{2',2}\Gamma^{\Lambda}\left( 1',2';1,2\right)S^{\Lambda}\left(2,2'\right), \\ \frac{d}{d\Lambda}\Gamma^{\Lambda}\left(1',2';1,2\right)&=T  \sum\limits_{3',3} \Gamma_3^\Lambda \left( 1',2',3';1,2,3\right) S^{\Lambda}\left(3,3'\right) \nonumber \\
 +T\sum\limits_{3',3;4',4}&\Bigg[ \Gamma^{\Lambda}\left( 1',2';3,4\right) \Gamma^{\Lambda}\left( 3',4';1,2\right) \nonumber \\ -\Gamma^{\Lambda}( 1',4'&;1,3) \Gamma^{\Lambda}\left( 3',2';4,2\right) - \left (3'\leftrightarrow 4', 3 \leftrightarrow 4 \right ) \nonumber \\ +\Gamma^{\Lambda}( 2',4'&;1,3) \Gamma^{\Lambda}\left( 3',1';4,2\right) + \left (3'\leftrightarrow 4', 3 \leftrightarrow 4 \right ) \Bigg ] \nonumber \\ \label{eq:SecondFlow} \times G^{\Lambda}(3,3'&)S^{\Lambda}(4,4'),
\end{align}
where $\Gamma_3$ is the three-particle vertex and $T$ denotes the system's temperature. All arguments ``$1$'' denote a collection of the Matsubara frequency, lattice site, and spin index, i.e., $1=\{\omega_1,i_1,\alpha_1\}$. The fully dressed propagator $G^\Lambda$ is given by $G^{\Lambda}=\left[ \left [ G^{\Lambda}_0 \right ]^{-1} -\Sigma^{\Lambda}\right ]^{-1}$ and $S^\Lambda$ denotes the so-called single-scale propagator
\begin{equation}
\label{eq:SingleScale}
 S^{\Lambda}=G^{\Lambda} \frac{d}{d\Lambda}\left[ G^{\Lambda}_0 \right ]^{-1} G^{\Lambda}\;,
\end{equation}
which occurs whenever the scale derivative acts on the free propagator. Below, these equations will be evaluated in the limit $T\rightarrow 0$ where the Matsubara sums become integrals with a prefactor $T\rightarrow\frac{d\omega}{2\pi}$. For a diagrammatic representation of Eqs.~\eqref{eq:FirstFlow} and \eqref{eq:SecondFlow}, see Fig.\ \ref{fig:FRGEquations}.

While the infinite set of FRG equations is formally exact, any numerical evaluation requires some type of truncation scheme. A numerically feasible scheme that has proven to correctly describe the magnetic properties of a wide class of spin systems is the so-called Katanin truncation~\cite{Katanin2004}. Within this approach, the contribution from the three-particle vertex $\Gamma_3^\Lambda$ in Eq.~\eqref{eq:SecondFlow} is neglected and the single-scale propagator $S^\Lambda$ is replaced by
\begin{equation}
 \label{eq:Katanin}
 S^{\Lambda}\longrightarrow -\frac{d}{d\Lambda}G^{\Lambda}=S^{\Lambda}-\left(G^{\Lambda}\right)^2\frac{d}{d\Lambda}\Sigma^{\Lambda}\;.
\end{equation}
Effectively, the replacement \eqref{eq:Katanin} is equivalent to the inclusion of a certain subset of three-particle vertices which are responsible for the feedback of the self energy into the flow of the two-particle vertex. It is important to stress that this feedback represents a significant advantage of the Katanin scheme compared to the -- seemingly more standard -- plain two-particle truncation without the replacement Eq. \eqref{eq:Katanin}. While the fully self-consistent Katanin scheme guarantees the {\it complete} feedback of $\Sigma^\Lambda$ into the flow of $\Gamma^\Lambda$, the plain two-particle truncation approximates this feedback at an intermediate level such that self-energy effects are insufficiently taken into account~\cite{Reuther2010}. As a consequence, a plain two-particle truncation cannot describe the formation of magnetically disordered phases and rather remains on a classical level of approximation~\cite{Reuther2010}. It has also been argued that the Katanin truncation leads to a better fulfillment of Ward identities associated with conservation laws~\cite{Katanin2004,Salmhofer2004}.

To numerically evaluate Eqs.~\eqref{eq:FirstFlow} and \eqref{eq:SecondFlow}, the frequency, site, and spin dependencies of the vertex functions need to be parameterized. We start with the self energy $\Sigma^{\Lambda}(1,2)$ which we rewrite as
\begin{equation}
 \Sigma^{\Lambda}(1,2)=-i\gamma_{\text{d}}^{\Lambda}(\omega_1)\delta_{i_1 i_2}\delta_{\alpha_1 \alpha_2}\delta(\omega_1 -\omega_2)\;.\label{selfenergy}
\end{equation}
The diagonal structures in frequencies and site indices are due to energy conservation and locality of the propagators, respectively. In the case of Heisenberg interactions, the self energy is also diagonal in spin space as expressed by the term $\delta_{\alpha_1 \alpha_2}$. Also note that the $SU(2)$ spin-rotation invariance dictates that the self energy is purely imaginary and antisymmetric in frequency, i.e., $\text{Im}\;\gamma_\text{d}^\Lambda(\omega)=0$ and $\gamma_\text{d}^\Lambda(\omega)=-\gamma_\text{d}^\Lambda(-\omega)$. The self energy, hence, accounts for a finite lifetime of the pseudo fermions. Furthermore, for lattices where all sites are {\it equivalent} (such as the kagome lattice), $\gamma_\text{d}^{\Lambda}(\omega)$ does not depend on the site. 

To implement spin-rotation symmetry for the fermionic two-particle vertex $\Gamma^{\Lambda}\left( 1',2';1,2\right)$, we note that (up to swapping indices) there are only two $4$-rank tensors in spin space that are invariant under $SU(2)$ transformations, $\sum_\mu\sigma^{\mu}_{\alpha_{1'} \alpha_{1}}\sigma^{\mu}_{\alpha_{2'} \alpha_{2}}$ and $\delta_{\alpha_{1'} \alpha_{1}}\delta_{\alpha_{2'} \alpha_{2}}$, representing the spin and density channel of the vertex, respectively. With these two terms we can parametrize the two-particle vertex by
\begin{align}\label{eq:AntySymmVertex}
 \Gamma^{\Lambda}\left( 1',2';1,2\right) & = \Bigg [ \Gamma^{\Lambda}_{\text{s} \, i_1 i_2}\left( \omega_1',\omega_2';\omega_1,\omega_2\right)\sum_\mu\sigma^{\mu}_{\alpha_{1'} \alpha_{1}}\sigma^{\mu}_{\alpha_{2'} \alpha_{2}} \nonumber \\ &+\Gamma^{\Lambda}_{\text{d} \, i_1 i_2}\left( \omega_1',\omega_2';\omega_1,\omega_2\right)\delta_{\alpha_{1'} \alpha_{1}}\delta_{\alpha_{2'} \alpha_{2}} \Bigg ] \nonumber \\ &\times \delta(\omega_1+\omega_2-\omega_{1'}-\omega_{2'})\delta_{i_{1'}i_1}\delta_{i_{2'}i_2} \nonumber \\ &- \left ( \omega_1\leftrightarrow \omega_2, \, \alpha_1 \leftrightarrow \alpha_2, \, i_1 \leftrightarrow i_2 \right ).
\end{align}
Here, the last line ensures that the vertex is fully antisymmetric under the exchange of $1\leftrightarrow 2$ or $1'\leftrightarrow 2'$, and the Kronecker deltas in real space are again a consequence of the bare propagator's locality. The flow equations can now be formulated in terms of the spin and density parts of the vertex, $\Gamma^{\Lambda}_{\text{s}}$ and $\Gamma^{\Lambda}_{\text{d}}$. Due to energy conservation, a description with three frequency arguments is sufficient and we can write
\begin{equation}
 \Gamma^{\Lambda}_{\text{s/d} \, i_1 i_2}\left( \omega_1',\omega_2';\omega_1,\omega_2\right)\longrightarrow \Gamma^{\Lambda}_{\text{s/d} \, i_1 i_2}\left( s,t,u\right),
\end{equation}
where the so-called transfer frequencies $s$, $t$, $u$ are given by $s=\omega_{1'}+\omega_{2'}$, $t=\omega_{1'}-\omega_{1}$, and $u=\omega_{1'}-\omega_{2}$. The explicit flow equations resulting from these parameterizations can be found in Ref.~\onlinecite{Reuther2010}.

For an efficient numerical solution, it is important to exploit all symmetries of the vertex functions in frequency and real-space variables. In particular, one can show that $\Gamma_\text{s}^\Lambda$ and $\Gamma_\text{d}^\Lambda$ fulfill the relations
\begin{subequations}
\begin{align} 
\Gamma^{\Lambda}_{\text{s/d} \, i_1 i_2}\left( s,t,u\right)&=\Gamma^{\Lambda}_{\text{s/d} \, i_2 i_1}\left( -s,t,u\right)\;,\label{symm1}\\
\Gamma^{\Lambda}_{\text{s/d} \, i_1 i_2}\left( s,t,u\right)&=\Gamma^{\Lambda}_{\text{s/d} \, i_1 i_2}\left( s,-t,u\right)\;,\label{symm2}\\
\Gamma^{\Lambda}_{\text{s/d} \, i_1 i_2}\left( s,t,u\right)&=\Gamma^{\Lambda}_{\text{s/d} \, i_2 i_1}\left( s,t,-u\right)\;,\label{symm3}\\
\Gamma^{\Lambda}_{\text{s/d} \, i_1 i_2}\left( s,t,u\right)&=\pm\Gamma^{\Lambda}_{\text{s/d} \, i_1 i_2}\left( u,t,s\right)\;,\label{symm4}
\end{align}
\end{subequations}
where in the last line the plus (minus) sign corresponds to the spin (density) channel. These properties lead to a reduction of the numerical effort by a factor of $2^4=16$.

The RG equations are solved with the initial conditions in the limit $\Lambda \rightarrow \infty$ given by the bare interactions, i.e., $\Gamma_{\text{d}\,i_1 i_2}^{\Lambda\rightarrow\infty}=0$ and $\Gamma_{\text{s}\,i_1 i_2}^{\Lambda\rightarrow\infty}=J_{i_1 i_2}/4$ [the factor of $1/4$ results from the fermionic representation in Eq.~\eqref{eq:AuxiliaryFermions}]. We evaluate the vertices at $50$ discrete frequency points for $s$, $t$, and $u$ which are chosen as a combination of a linear and logarithmic mesh. The real-space dependence of $\Gamma_{\text{s/d}\,i_1 i_2}$ is approximated by neglecting all vertices where the distance between $i_1$ and $i_2$ exceeds a maximal value. Here, the maximal distance is chosen to be seven nearest-neighbor lattice spacings which means that correlations are considered within a hexagon of $131$ lattice sites. 

Connecting the pairs of external legs $(1,1')$ and $(2,2')$ of the two-particle vertex $\Gamma^{\Lambda}\left( 1',2';1,2\right)$ and integrating over the corresponding frequencies, directly yields the spin-spin correlator defined for imaginary frequencies $i\Omega$,
\begin{equation}
 \chi_{ij}^{\mu\nu}(i\Omega)=\int_0^{\infty} d\tau e^{i\Omega \tau}\left < T_{\tau}S^{\mu}_i(\tau)S^{\nu}_j(0) \right >\;,\label{correlator}
\end{equation}
where $\tau$ is an imaginary-time variable. The central quantity to be studied within PFFRG is the $\vec{k}$-space-resolved static susceptibility $\chi^{\mu\nu,\Lambda}(\vec{k})\equiv\chi^{\mu\nu,\Lambda}(\vec{k},i\Omega=0)$ given by the $\Omega=0$ component of the Fourier transform of Eq.~\eqref{correlator} evaluated as a function of $\Lambda$. Most importantly, the $\Lambda$ behavior of the susceptibility contains information about the magnetic properties of the system. If magnetic \textit{long-range} order sets in, a divergence of the susceptibility and a breakdown of the $\Lambda$-dependent flow is expected. This is explained by the fact that our PFFRG scheme does strictly not allow for spontaneous symmetry breaking. In a finite system with discretized frequencies, such a divergence is regularized and manifests as a kink or a cusp as $\Lambda$ is decreased. The point in $\vec{k}$ space at which this anomaly occurs further indicates the wave vector of the corresponding type of magnetic order. On the other hand, a smooth $\Lambda$ flow of the susceptibility down to the physical limit $\Lambda=0$ signals the absence of any type of magnetic long-range order and indicates a magnetically disordered ground state. In this case, the momentum-space profile of $\chi^{\mu\nu,\Lambda=0}(\vec{k})$ still allows to identify the wave vectors of the dominant {\it short-range} spin-spin correlations.

\subsection{Modifications of the PFFRG for finite DM interactions}
\label{sec:FRGB}
The central modification of the PFFRG approach in the case of finite DM interactions concerns the parameterization of the vertices. Since the DM exchange breaks the $SU(2)$ spin symmetry down to $U(1)$ rotations around the $z$ axis, spin terms of the form $\sum_{\mu=x,y}\sigma^{\mu}_{\alpha_{1'} \alpha_{1}}\sigma^{\mu}_{\alpha_{2'} \alpha_{2}}$ and $\sigma^z_{\alpha_{1'} \alpha_{1}}\sigma^z_{\alpha_{2'} \alpha_{2}}$ need to be parameterized by two distinct two-particle vertices. As a consequence, the spin vertex $\Gamma_\text{s}^\Lambda$ is replaced by two vertices $\Gamma_{xx}^\Lambda=\Gamma_{yy}^\Lambda$ and $\Gamma_{zz}^\Lambda$. Together with $\Gamma^{\Lambda}_{d}$, these three vertices are sufficient to treat models with $XXZ$-interactions as shown, e.g., in Ref.\ \citenum{Iqbal2016a}. The case of DM interactions is, however, more involved. To implement the spin structure $\sigma^x_{\alpha_{1'} \alpha_{1}}\sigma^y_{\alpha_{2'} \alpha_{2}}-\sigma^y_{\alpha_{1'} \alpha_{1}}\sigma^x_{\alpha_{2'} \alpha_{2}}$ of the DM interaction, a vertex $\Gamma_{\text{DM}}^\Lambda$ needs to be introduced. At the initial value $\Lambda\rightarrow\infty$, this vertex is given by the bare DM coupling $\Gamma_{\text{DM}\,i_1 i_2}^{\Lambda\rightarrow\infty}=D_{i_1 i_2}/4$. Additionally, two more distinct vertices $\Gamma^{\Lambda}_{z\text{d}}$ and $\Gamma^{\Lambda}_{\text{d}z}$ parameterizing the $U(1)$-invariant spin terms $\sigma^z_{\alpha_{1'} \alpha_{1}}\delta_{\alpha_{2'}\alpha_{2}}$ and $\delta_{\alpha_{1'}\alpha_{1}}\sigma^z_{\alpha_{2'} \alpha_{2}}$ must be considered. The full set of $U(1)$-symmetric two-particle vertices is hence given by $\Gamma_{xx}^\Lambda$, $\Gamma_{zz}^\Lambda$, $\Gamma_\text{d}^\Lambda$, $\Gamma_{\text{DM}}^\Lambda$, $\Gamma^{\Lambda}_{z\text{d}}$, and $\Gamma^{\Lambda}_{\text{d}z}$ and the parameterization of $\Gamma^{\Lambda}\left( 1',2';1,2\right)$ reads
\begin{widetext}
\begin{align}
 \Gamma^{\Lambda}\left( 1',2';1,2\right) &= \Bigg [ \Gamma^{\Lambda}_{xx \, i_1 i_2}\left( s,t,u\right)\left(\sigma^{x}_{\alpha_{1'} \alpha_{1}}\sigma^{x}_{\alpha_{2'} \alpha_{2}} + \sigma^{y}_{\alpha_{1'} \alpha_{1}}\sigma^{y}_{\alpha_{2'} \alpha_{2}}\right) + \Gamma^{\Lambda}_{zz \, i_1 i_2}\left( s,t,u\right) \sigma^{z}_{\alpha_{1'} \alpha_{1}}\sigma^{z}_{\alpha_{2'} \alpha_{2}} \nonumber \\ &+\Gamma^{\Lambda}_{\text{DM} \, i_1 i_2}\left( s,t,u\right)\left(\sigma^{x}_{\alpha_{1'} \alpha_{1}}\sigma^{y}_{\alpha_{2'} \alpha_{2}} - \sigma^{y}_{\alpha_{1'} \alpha_{1}}\sigma^{x}_{\alpha_{2'} \alpha_{2}}\right) + \Gamma^{\Lambda}_{\text{d} \, i_1 i_2}\left( s,t,u\right)\delta_{\alpha_{1'} \alpha_{1}}\delta_{\alpha_{2'} \alpha_{2}} \nonumber \\ &+ \Gamma^{\Lambda}_{z\text{d} \, i_1 i_2}\left( s,t,u\right)\sigma^{z}_{\alpha_{1'} \alpha_{1}}\delta_{\alpha_{2'} \alpha_{2}} + \Gamma^{\Lambda}_{\text{d}z \, i_1 i_2}\left( s,t,u\right)\delta_{\alpha_{1'} \alpha_{1}}\sigma^{z}_{\alpha_{2'} \alpha_{2}} \Bigg ] \delta(\omega_1+\omega_2-\omega_{1'}-\omega_{2'})\delta_{i_{1'}i_1}\delta_{i_{2'}i_2} \nonumber \\ &- \left ( \omega_1\leftrightarrow \omega_2, \, \alpha_1 \leftrightarrow \alpha_2, \, i_1 \leftrightarrow i_2 \right )\;.\label{parametrize_DM}
\end{align}
\end{widetext}
Since the two-particle vertex couples to the flow of the self energy, the parameterization of $\Sigma^{\Lambda}(1,2)$ is also modified. In addition to the density term $\gamma_\text{d}^\Lambda$ in Eq.~\eqref{selfenergy}, a spin-dependent term $\gamma_\text{s}^\Lambda$ is generated during the RG flow such that the full parameterization of the self energy is given by
\begin{equation}
 \Sigma^{\Lambda}(1,2)=(-i\gamma_\text{d}^\Lambda(\omega_1)\delta_{\alpha_1 \alpha_2}+\gamma_\text{s}^\Lambda(\omega_1)\sigma^z_{\alpha_1 \alpha_2})\delta_{i_1 i_2}\delta(\omega_1 -\omega_2)\;,
\end{equation}
where $\gamma^\Lambda_\text{s}(\omega)$ is real and antisymmetric in its frequency argument. Even though the new self-energy term $\sim\gamma_\text{s}^\Lambda(\omega)\sigma^z_{\alpha_1 \alpha_2}$ might appear to have the same form as an external magnetic field acting on the fermion system, this contribution does indeed not break time-reversal symmetry due to the property $\gamma_\text{s}(\omega)=-\gamma_\text{s}(-\omega)$ (a magnetic field, in contrast, would be symmetric in $\omega$). We note that $\gamma_\text{s}(\omega)$ solely appears due to the DM interaction and cannot be generated by diagonal types of exchange couplings.

The new parameterization of the self energy is also carried over to the dressed propagator $G^\Lambda$ and the single scale propagator $S^\Lambda$ which both acquire a spin part in addition to the density channel. In total, this complicates the RG equations enormously. First, all six two-particle vertices become finite during the flow and their contributions cannot be neglected. Furthermore, when inserting the parameterizations of the vertices into Eq.~\eqref{eq:SecondFlow}, the products $\Gamma^\Lambda\Gamma^\Lambda G^\Lambda S^\Lambda$ on the right-hand side of the equation generate all different types of terms containing combinations of the six two-particle vertices as well as the spin and density channels of $G^\Lambda$ and $S^\Lambda$. Finally, from the four symmetry relations in Eqs.~\eqref{symm1}-\eqref{symm4} only Eq.~\eqref{symm1} and a combination of Eqs.~\eqref{symm2} and \eqref{symm3} (which amounts to replacing $1\leftrightarrow2$ and $1'\leftrightarrow2'$) remain intact, resulting in an additional factor of four in the computation time. Together with the larger number of vertex functions, the computational effort due to finite DM interactions increases by a factor of $80$. Given the complexity of the RG equations, we will not write down their explicit form here, but continue discussing their solution for the kagome lattice in the next sections.

\section{$J_1$-$D$-model on the kagome lattice}
\label{sec:NearestNeighbor}

Before we turn to the more complex $J_1$-$J_2$-$D$ model on the kagome lattice, we consider the simpler nearest-neighbor model which results from Eq.\ \eqref{eq:NNNHamiltonian} by setting $J_2=0$ and $J_1>0$. Particularly, we benchmark our PFFRG results against other approaches to test whether this technique correctly describes the transition into the $\mathbf{q}=0$ ordered state.
\begin{figure}[t]
\includegraphics[width=3.0in]{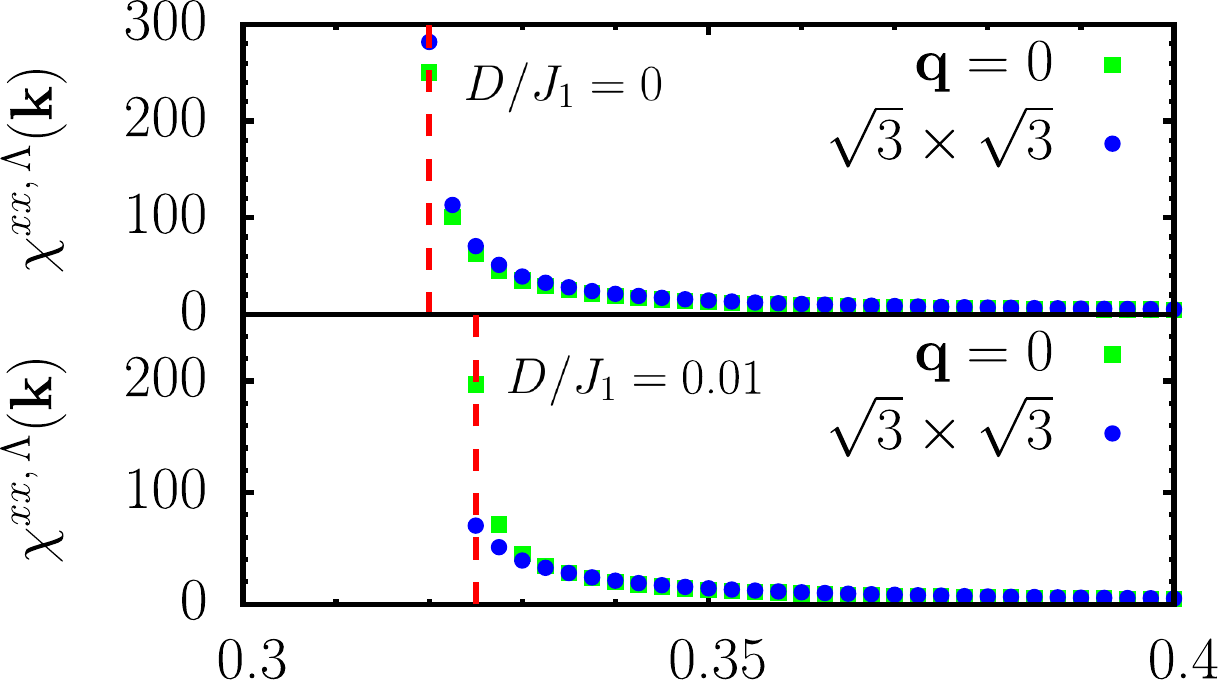}
\caption{(Color online) Flowing susceptibilities of the $J_1$-$D$ model treated within an RPA scheme: Shown are the susceptibilities for $\sqrt{3}\times \sqrt{3}$ order (blue circles) and $\vec{q}=0$ order (green squares). Upper panel: At vanishing DM coupling, $\sqrt{3}\times \sqrt{3}$ order is found to be preferred over $\vec{q}=0$ order (dashed red lines illustrate the critical $\Lambda$ scale at which the flow diverges first). Lower panel: At $D=0.01J_1$, the $\vec{q}=0$ order is dominant indicating a change of magnetization at infinitesimal $D$. For the peak positions of these types of order in reciprocal space, see Fig.\ \ref{fig:NNSus}(a). \label{fig:RPAJ1DFlow}}
\end{figure}

\subsection{PFFRG in the RPA channel}\label{sec:RPA}

As a first check, we verify that the new implementation of the PFFRG generally identifies the classical $\vec{q}=0$ order generated by the DM interaction. To this end, we analytically solve a simplified version of the PFFRG equations where only the RPA channel [also referred to as direct particle-hole channel, see third line of Eq.\ \eqref{eq:SecondFlow}] contributes to the flow of $\Gamma^{\Lambda}$ and self-energy effects are neglected. The flow equation for the two-particle vertex then reduces to
\begin{align}
 \frac{d}{d\Lambda}\Gamma^{\Lambda}\left(1',2';1,2\right)&=-T\hspace*{-6pt}\sum\limits_{3',3;4',4} \hspace*{-6pt}\left[\Gamma^{\Lambda}( 1',4';1,3) \Gamma^{\Lambda}\left( 3',2';4,2\right)\right.\notag\\
 &+ \left.\left (3'\leftrightarrow 4', 3 \leftrightarrow 4 \right )\right] G^{\Lambda}(3,3')S^{\Lambda}(4,4')\;.\label{RPA}
\end{align}
Singling out this channel is equivalent to treating a large $S$ generalization of the spin model (where $S$ is the spin length) and allows us to determine the type of classical order the system tends to establish in this limit~\cite{Baez2016}. It is worth noting that, due to its special real-space structure, the RPA channel is the only term in the PFFRG equations that generates long-range correlations between spins.

A PFFRG scheme in the RPA channel leads to substantial simplifications. Inserting the parameterization of Eq.~\eqref{parametrize_DM} into Eq.~\eqref{RPA}, one finds that $\Gamma^\Lambda_{\text{d}}$, $\Gamma^\Lambda_{z\text{d}}$, and $\Gamma^\Lambda_{\text{d}z}$ remain exactly zero during the entire RG flow. Furthermore, two-particle vertices at different frequency grid points decouple such that we can restrict ourselves to the zero frequency component. The resulting set of equations for the static ($s=t=u=0$) two-particle vertices reads
\begin{subequations}\label{eq:RPAfirst}
 \begin{align}
  \frac{d}{d\Lambda}\Gamma^{\Lambda}_{xx \, i_1 i_2} &= \frac{2}{\pi\Lambda^2}\sum\limits_{j}\left (\Gamma^{\Lambda}_{xx \, i_1j} \Gamma^{\Lambda}_{xx \, ji_2} - \Gamma^{\Lambda}_{\text{DM} \, i_1j} \Gamma^{\Lambda}_{\text{DM} \, ji_2} \right ), \\ \frac{d}{d\Lambda}\Gamma^{\Lambda}_{\text{DM} \, i_1i_2} &= \frac{2}{\pi\Lambda^2}\sum\limits_{j}\left (\Gamma^{\Lambda}_{\text{DM} \, i_1j} \Gamma^{\Lambda}_{xx \, ji_2} +\Gamma^{\Lambda}_{xx \, i_1j} \Gamma^{\Lambda}_{\text{DM} \, ji_2} \right ),\\
  \frac{d}{d\Lambda}\Gamma^{\Lambda}_{zz \, i_1 i_2} &= \frac{2}{\pi\Lambda^2}\sum\limits_{j}\Gamma^{\Lambda}_{zz \, i_1j} \Gamma^{\Lambda}_{zz \, ji_2}\;.
 \end{align}
\end{subequations}
One can see that the DM vertex $\Gamma^{\Lambda}_{\text{DM}}$ only couples to $\Gamma^{\Lambda}_{xx}$ (and vice versa) while $\Gamma^{\Lambda}_{zz}$ is completely unaffected by the DM interaction. Since at finite DM couplings $\Gamma^{\Lambda}_{zz}$ is generally found to be smaller than $\Gamma^{\Lambda}_{xx}$ (which is equivalent to the statement that spins favor an orientation in the $x$-$y$ plane), only $\Gamma^{\Lambda}_{xx}$ and $\Gamma^{\Lambda}_{\text{DM}}$ are considered below.

In the next step, we Fourier-transform the vertices via
\begin{equation}
\Gamma^\Lambda_{xx/\text{DM}\,a(i)b(j)}(\vec{k})=\sum_{\Delta \vec{R}=\vec{R}_i-\vec{R}_j}e^{-i\vec{k}(\vec{R}_i-\vec{R}_j)}\Gamma^\Lambda_{xx/\text{DM}\,ij}\;.
\end{equation}
Here, $a(i)=1,2,3$ is the sublattice index of site $i$ and the same holds for $b(j)$. $\vec{R}_i$ denotes the position of the kagome-unit cell in which site $i$ resides. With this transformation, the two-particle vertices become $3\times3$ matrices in the sublattice index and different Fourier components in $\vec{k}$ space decouple. The RG equations can be further decoupled with respect to the $xx$ and DM channels by defining vertices $\Gamma^{\Lambda}_{\pm\,ab}(\vec{k})=\Gamma^{\Lambda}_{xx\,ab}(\vec{k})\pm i\Gamma^{\Lambda}_{\text{DM}\,ab}(\vec{k})$, yielding
\begin{equation}
  \frac{d}{d\Lambda}\Gamma^{\Lambda}_{\pm} (\vec{k})= \frac{2}{\pi\Lambda^2} \Gamma^{\Lambda}_{\pm}(\vec{k})\Gamma^{\Lambda}_{\pm}(\vec{k})\;.
\end{equation}
Since the product of vertices on the right hand side is a standard matrix product in the sublattice indices, we have suppressed all sublattice variables. The solution of this equation is given by
 \begin{equation}
  \Gamma^{\Lambda}_{\pm} (\vec{k})=\pi\Lambda\left[2+\pi\Lambda \left( \Gamma^{\Lambda\rightarrow\infty}_{xx}(\vec{k})\pm i\Gamma^{\Lambda\rightarrow\infty}_{\text{DM}}(\vec{k}) \right)^{-1}\right]^{-1}\;.
 \end{equation}
Transforming back to the original vertices $\Gamma^{\Lambda}_{xx}$ and $\Gamma^{\Lambda}_{\text{DM}}$, we obtain the spin susceptibility $\chi^{xx,\Lambda}_{\vec{k}}=\chi^{yy,\Lambda}_{\vec{k}}$ as a function of $\Lambda$ which we use to probe the magnetic order in the $x$-$y$ plane.

\begin{figure*}[t]
\includegraphics[width=6.8in]{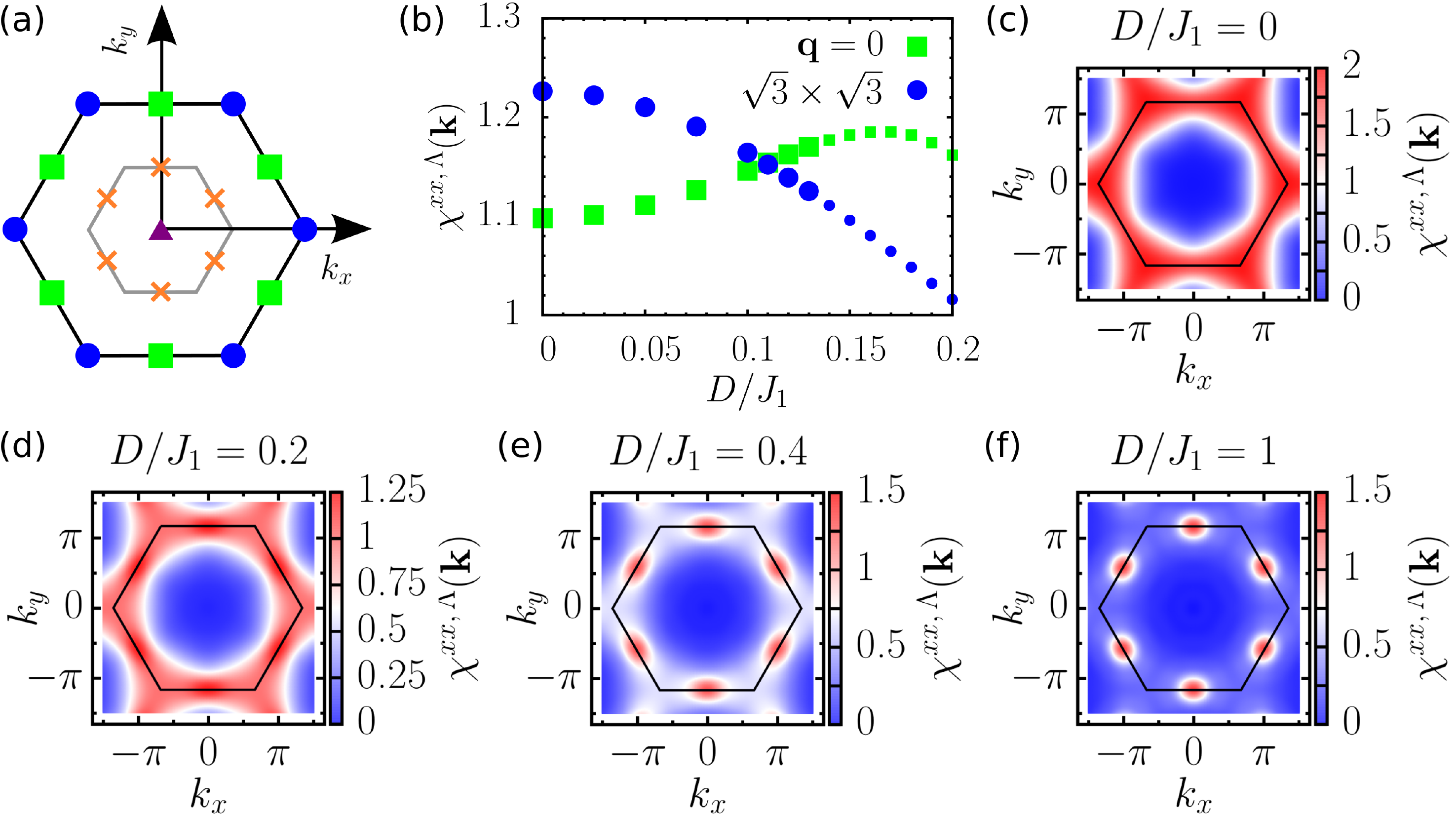}
\caption{(Color online) (a) Dominant peak positions for the four types of classical order in the $J_1$-$J_2$ kagome Heisenberg model: Ferromagnetic order (purple triangle), $\sqrt{3}\times\sqrt{3}$ order (blue circles), $\vec{q}=0$ order (green squares), and cuboc order (orange crosses) are shown within the boundaries of the extended Brillouin zone (black hexagon). The edges of the first Brillouin zone are depicted gray. (b) Competition between the $\vec{q}=0$ (green squares) and $\sqrt{3}\times\sqrt{3}$ (blue circles) susceptibilities as a function of $D$. The data corresponds to $\Lambda \approx 0.19$. Small symbols indicate that the flow has entered the symmetry broken regime below the critical $\Lambda$ where the $SU(2)$-invariant PFFRG approach is no longer valid. (c)-(f) Static spin susceptibilities $\chi^{xx,\Lambda}(\vec{k})$ for various DM interaction strengths. The black hexagon denotes the boundaries of the extended Brillouin zone. Note that, in magnetically disordered regimes such as (c), the plot corresponds to $\Lambda=0$ while in (d), (e), and (f) the susceptibility is shown at a $\Lambda$ value right above the $\vec{q}=0$ instability (indicated by arrows in Fig.~\ref{fig:J1DFLow}). \label{fig:NNSus}}
\end{figure*}

Due to the classical nature of the RG equations in the RPA channel, the susceptibilities always diverge during the $\Lambda$ flow indicating the onset of magnetic order. To identify the type of order that is classically preferred, we determine the wave vector $\vec{k}$ for which this divergence occurs {\it first} as $\Lambda$ is lowered. Generally, the competition of different orders only takes place between $\sqrt{3}\times\sqrt{3}$ and $\vec{q}=0$ states [see Fig.~\ref{fig:NNSus}(a) for the corresponding wave-vector positions in reciprocal space]. At vanishing DM coupling, we find that the $\sqrt{3}\times\sqrt{3}$ susceptibility slightly dominates over the $\vec{q}=0$ susceptibility, see Fig.\ \ref{fig:RPAJ1DFlow}. This is consistent with earlier semi-classical studies of the model~\cite{Chubukov1992,Chernyshev2014} which predict a preference for $\sqrt{3}\times\sqrt{3}$ order at large $S$. Switching on an infinitesimal DM interaction $D>0$ the situation is found to be reversed: The $\vec{q}=0$ susceptibility diverges at slightly larger $\Lambda$ as compared to the $\sqrt{3}\times\sqrt{3}$ component, suggesting that the system now realizes $\vec{q}=0$ order. Together with the observation that the $\vec{q}=0$ phase persists up to $D\rightarrow\infty$, this is exactly the semi-classical result of Ref.~\onlinecite{Elhajal2002}.

Overall, this simplified PFFRG approach shows that $\vec{q}=0$ order is correctly selected at finite DM interactions. The absence of any non-magnetic phases and the onset of $\vec{q}=0$ order at infinitesimally small $D$ is, of course, an artifact of the classical treatment. In the next section, we investigate how quantum fluctuations change this picture.

\begin{figure}[t]
\includegraphics[width=3.0in]{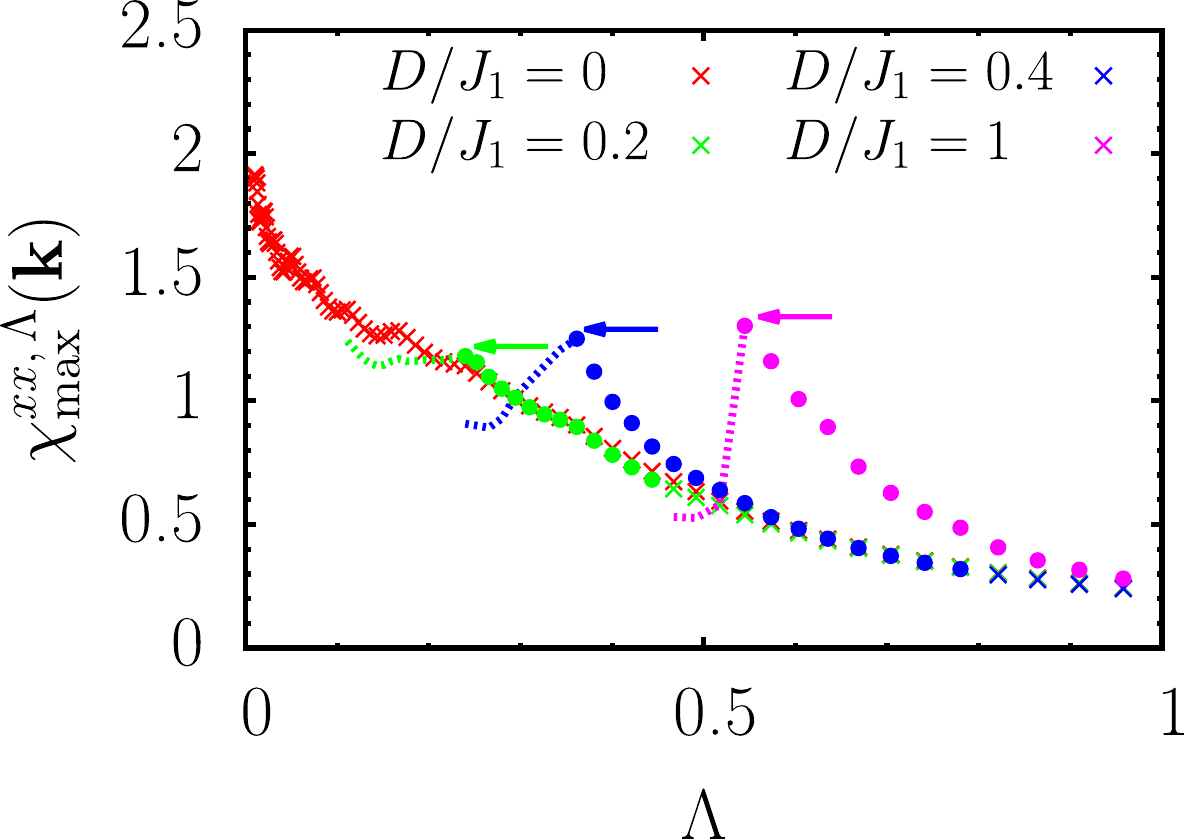}
\caption{(Color online) $\Lambda$ dependence of the static susceptibility $\chi^{xx,\Lambda}(\vec{k})$ for various DM-interaction strengths taken at the maximum in momentum space. Crosses (circles) indicate that the maximum in $\vec{k}$ space at the respective $\Lambda$ and $D$ values is located at the $\sqrt{3}\times\sqrt{3}$ ($\vec{q}=0$) position. Instability features associated with critical $\Lambda$ scales are marked by arrows. Below those scales, the susceptibilities are plotted by dashed lines. \label{fig:J1DFLow}}
\end{figure}

\subsection{Full PFFRG calculation}

We now discuss the results of a full quantum-PFFRG calculation taking into account all interaction channels of Eq.\ \eqref{eq:SecondFlow}. As in the previous section, the effects of the DM interaction are most pronounced in the $x$-$y$ plane such that we restrict our discussion to in-plane susceptibilities. In Fig.\ \ref{fig:NNSus}(c)-(f), we show $\vec{k}$-space-resolved susceptibility profiles $\chi^{xx,\Lambda}(\vec{k})$ for selected values of the DM interaction. For vanishing DM coupling [see \ref{fig:NNSus}(c)], we reproduce the profile that has previously been obtained by PFFRG~\cite{Suttner2014} showing the strongest signal at the boundaries of the extended Brillouin zone and small maxima at the $\sqrt{3}\times\sqrt{3}$ positions. At the same time, the flow does not display signs of an instability (see Fig.\ \ref{fig:J1DFLow}) which hints at a non-magnetic ground state. Comparing our results with other numerical methods, there is broad consensus that the response in momentum space is mostly distributed along the edge of the extended Brillouin zone~\cite{Laeuchli2009,Depenbrock2012,Shimokawa2016}. However, the position of the residual small peaks is still debated: While DMRG studies on tori find the $\sqrt{3}\times\sqrt{3}$ positions preferred~\cite{Depenbrock2012}, exact diagonalization of small spin clusters detects maxima at $\vec{q}=0$ positions for $T=0$\cite{Laeuchli2009,Shimokawa2016}.

When $D$ is increased, the response first remains rather evenly distributed along the Brillouin-zone edges, but the small peaks shift towards the $\vec{q}=0$ positions, see Fig.~\ref{fig:NNSus}(d). Only as $D$ is increased beyond $D\gtrsim 0.4J_1$, the $\vec{q}=0$ peaks become more prominent and the ridge-like feature along the Brillouin-zone boundary disappears, see Fig.~\ref{fig:NNSus}(e),(f). To investigate this change in more detail, we compare the susceptibilities for $\sqrt{3}\times\sqrt{3}$ and $\vec{q}=0$ orders as a function of $D$ in Fig.~\ref{fig:NNSus}(b). One can see that the point at which the $\vec{q}=0$ susceptibility surpasses the $\sqrt{3}\times\sqrt{3}$ response is rather exactly given by $D=0.11J_1$ [note that the data in Fig.~\ref{fig:NNSus}(b) corresponds to finite $\Lambda\approx 0.19$ which is also the value where a $\vec{q}=0$ instability is observed, see below].

While our results indicate that the magnetic correlations undergo a qualitative change at $D=0.11 J_1$, we need to detect signatures of an instability during the RG flow to confirm that this change is associated with the onset of magnetic long-range order. In Fig.\ \ref{fig:J1DFLow}, we plot the flow behavior of the susceptibility for various values of $D/J_1$. While a kink at finite $\Lambda$ is clearly resolved for $D/J_1\gtrsim0.2$, determining the precise value for the critical DM-interaction strength turns out to be rather challenging. The reason for this is that in comparison to recent PFFRG studies for Heisenberg models, we use relatively small system sizes and coarse frequency grids which increases numerical oscillations due to frequency discretization. Additionally, the phase transition between the non-magnetic phase and the $\vec{q}=0$ phase appears to be rather smooth with a slow onset of magnetization. Our best estimate for the first appearance of an instability feature is $D=\left(0.12\pm0.02\right)J_1$ which also coincides with the rise of $\vec{q}=0$ peaks.

Taken together, the change of spin correlations in conjunction with the onset of instability signatures at $D\approx 0.1J_1$ indicates that PFFRG correctly reproduces the phase diagram of the $J_1$-$D$ model that has previously been obtained by exact diagonalization~\cite{Cepas2008}. We also conclude that PFFRG incorporates the proper amount of quantum fluctuations to balance between magnetic order and disorder tendencies. We therefore continue exploring more complex models within this formalism in the next section.

\section{$J_1$-$J_2$-$D$ model on the kagome lattice}\label{sec:NextNearestNeighbor}
\subsection{Phase diagram}\label{sec:NextNearestNeighborA}

\begin{table}[t]
\begin{tabular}{c c c c c  }
\hline\hline
order type &$\vec{q}=0$  & cuboc & ferro & $\sqrt{3}\times\sqrt{3}$ \\
\hline
$D=0.0$ &$[27, 59]$ &$[122 , 153]$ & $[171,270]$& $[270,347]$\\
$D=0.2$ &$[0, 81]$ & $[122 , 158]$&$[171,270]$ & $[270,347]$\\
$D=0.4$ &$[-6 , 95]$ & $[117 , 162]$& $[171,270]$& $[270,353]$\\
\hline\hline
\end{tabular}\caption{Phase boundaries for the ordered phases of the $J_1$-$J_2$-$D$ model on the kagome lattice as found via PFFRG. The Heisenberg couplings are parameterized as $J_1 = J\cos{\theta}$, $J_2 = J\sin{\theta}$ and the $\theta$ intervals are given in angular degrees. The accuracy of the $\theta$ values is roughly $\pm 5^{\circ}$.}\label{tab:OrderingAngles}
\end{table}

Let us now consider the full Hamiltonian in Eq.\ \eqref{eq:NNNHamiltonian} and investigate the resulting phase diagram for positive and negative Heisenberg couplings. As discussed in the next section, the case of dominant $J_1>0$ and smaller $D$, $J_2>0$ is relevant for \textit{herbertsmithite}. The Heisenberg interactions are parameterized by an angle $\theta$ and an overall amplitude $J$, i.e., we set $J_1 = J\cos{\theta}$, $J_2 = J\sin{\theta}$.

\begin{figure*}[t]
\includegraphics[width=6.0in]{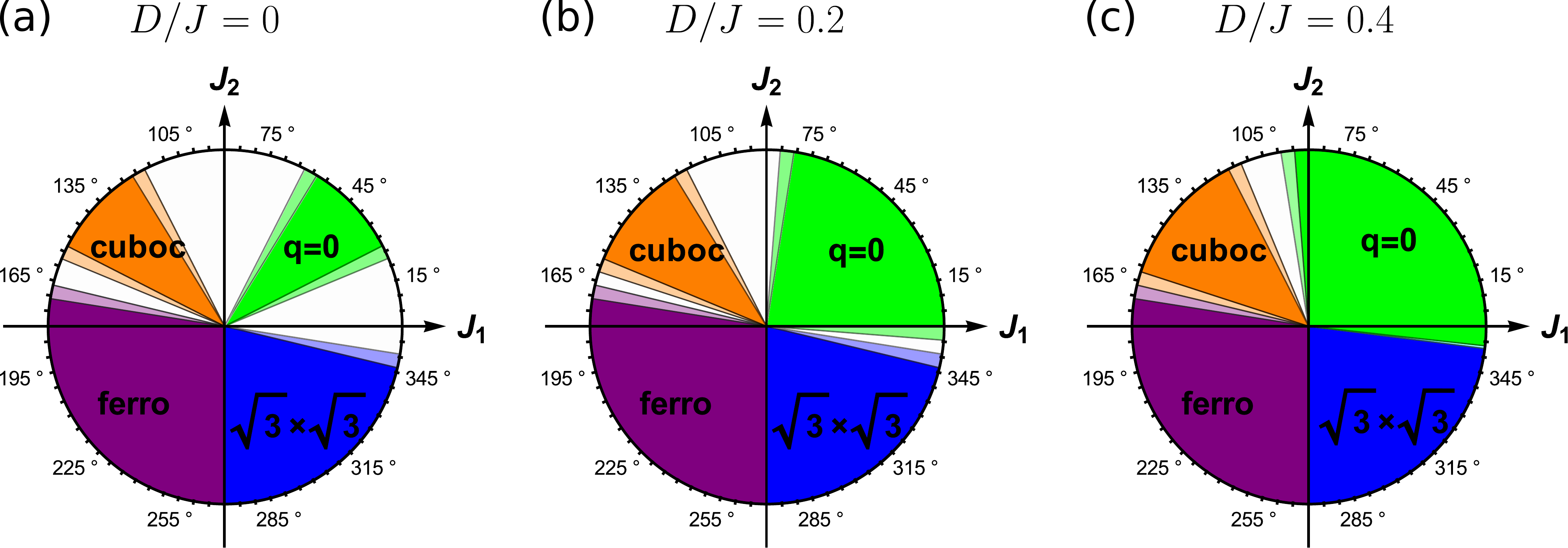}
\caption{(Color online) Phase diagram of the $J_1$-$J_2$-$D$ model as a function of $\theta\in[0,2\pi)$ and various values of $D$: Colored regions are the classically ordered phases of Fig.~\ref{fig:DMkagome}(b) while the white regimes are magnetically disordered. Uncertainties in the phase boundaries between magnetically ordered and non-magnetic phases are indicated by light-colored stripes. The $\theta$ values of the phase boundaries are also listed in Table~\ref{tab:OrderingAngles}.\label{fig:PhaseDiagrams}}
\end{figure*}

Without the DM interaction, this model has already been studied with PFFRG~\cite{Suttner2014, Buessen2016}. In agreement with these works, we obtain all types of order of the classical phase diagram, but with additional non-magnetic phases opening up around the points $(J_1,J_2) = (1,0)$ and $(0,1)$, see Fig.~\ref{fig:PhaseDiagrams}(a). Further, our results indicate the possible existence of a narrow non-magnetic phase between ferromagnetic and cuboc regimes. Compared to Ref.\ \citenum{Buessen2016}, the magnetically disordered phases are found to be slightly larger, possibly because we use smaller system sizes and fewer discrete frequencies which complicates the identification of magnetic instabilities.

A finite DM interaction first has the biggest effect on the $\vec{q}=0$ phase which is considerably enlarged upon increasing $D$. At $D=0.2J$ [see Fig.~\ref{fig:PhaseDiagrams}(b)], the $\vec{q}=0$ regime almost fills the whole first quadrant of the phase diagram and the non-magnetic phases around $(J_1,J_2) = (1,0)$ and $(0,1)$ shrink, accordingly. Further increasing $D$ [Fig.~\ref{fig:PhaseDiagrams}(c)], we even find $\vec{q}=0$ order for ferromagnetic couplings $J_1<0$ or $J_2<0$ and the $\sqrt{3}\times\sqrt{3}$ and cuboc phases likewise undergo enlargements. As a consequence, the non-magnetic phase around $\theta=0$ has completely vanished at $D=0.4J$. Note that, for all DM couplings which we have studied, the transition between ferromagnetic and $\sqrt{3}\times\sqrt{3}$ phases remains exactly at $\theta=3\pi/2$ (negative $J_2$ axis). The precise $\theta$ intervals for the ordered phases are listed in Table~\ref{tab:OrderingAngles}.

In summary, these results show that in parameter regions where a non-collinear magnetic phase ($\vec{q}=0$, $\sqrt{3}\times\sqrt{3}$, or cuboc order) competes with a magnetically disordered regime, a finite DM interaction shifts the phase boundary in favor of the non-collinear state. This behavior is plausible since the DM coupling tends to induce finite angles between neighboring spins (the largest energy gain for two DM-coupled spins is obtained for an angle of $\pi/2$ between them) which generally promotes non-collinear types of order. In contrast, the ferromagnetic regime is found to remain unchanged upon increasing $D$. Our results further indicate that for strong enough DM couplings, non-magnetic phases die out completely on the kagome lattice.

\subsection{Implications for \textit{herbertsmithite}}\label{sec:NextNearestNeighborB}

Our results for the $J_1$-$J_2$-$D$ model on the kagome lattice can be used to study the microscopic coupling scenario of the mineral \textit{herbertsmithite} ($\mathrm{ZnCu_3(OH)_6 Cl_2}$). The immense interest in this material mainly stems from the fact that it does not exhibit signatures of magnetic long-range down to $50 \,mK$~\cite{Shores2005, Mendels2007, Helton2007} but shows a diffuse, spinon-like excitation spectrum~\cite{Han2012,Han2016}. \textit{Herbertsmithite}, hence, displays all the experimental features expected from a quantum spin liquid. The spin structure factor measured with neutron scattering features the strongest signal along the edges of the extended Brillouin zone which roughly resembles the momentum profile for a nearest-neighbor antiferromagnetic Heisenberg model on the kagome lattice. While early single-crystal neutron-scattering data did not resolve any preferred type of spin correlations along the edge~\cite{Han2012}, more recent results show small peaks at the $\vec{q}=0$ position~\cite{Han2016}. More insights into the microscopic couplings come from ESR measurements, magnetic susceptibility fittings, and the entropy difference compared to the Heisenberg case which indicate a DM interaction in the range of $D/J_1 \sim 0.08 \dots 0.1$ \cite{Rigol2007,Zorko2008,Singh2009}. In addition, ab-initio DFT calculations predict an antiferromagnetic second-neighbor interaction given by $J_2/J_1=0.019$\cite{Jeschke2013}.

\begin{figure*}[t]
\includegraphics[width=6.5in]{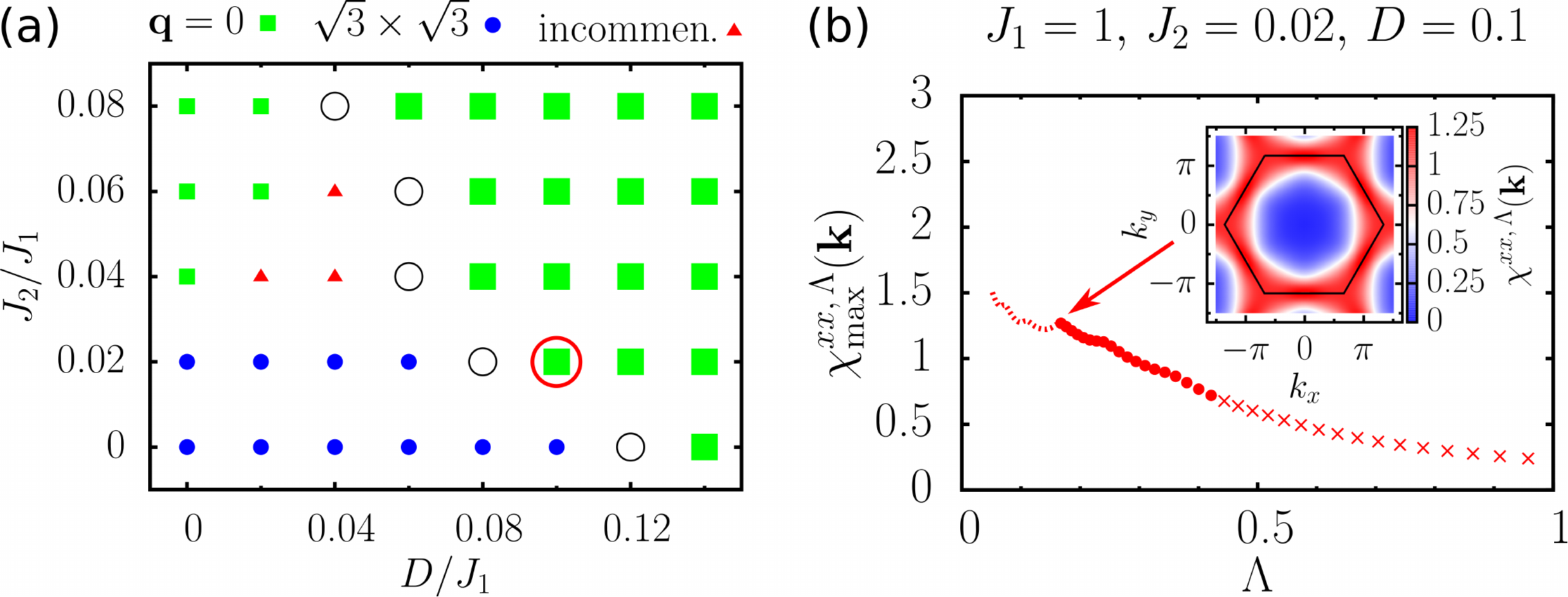}
\caption{(Color online) (a) Section of the phase diagram in the $J_2$-$D$ plane relevant for {\it herbertsmithite}: Large (small) icons denote that the RG flow does (does not) signal long-range order. Black circles indicate regions of numerical uncertainties where we cannot reliably determine the magnetic properties. The ordered phase is dominated by $\vec{q}=0$ order [confer Fig.~\ref{fig:NNSus} (a)], whereas, in the paramagnetic regime, we find dominant $\vec{q}=0$ (green squares), $\sqrt{3}\times\sqrt{3}$ (blue circles), as well as incommensurate (red triangles) spin fluctuations. (b) RG flow of the maximal susceptibility for $(J_1,J_2,D)=(1,0.02,0.1)$ [see red circle in (a)] showing a weak instability feature (arrow): As in Fig.~\ref{fig:J1DFLow}, crosses (circles) indicate that the maximum in momentum space resides at the $\sqrt{3}\times\sqrt{3}$ ($\vec{q}=0$) position. The inset shows the susceptibility in $\vec{k}$ space. Small maxima at the midpoints of the extended Brillouin zone's edges (corresponding to $\vec{q}=0$ correlations) are in agreement with low-energy inelastic neutron-scattering data\cite{Han2016}.  \label{fig:HerbertSus}}
\end{figure*}

We have performed PFFRG calculations in the vicinity of the reported values for $D$ and $J_2$, see Fig.~\ref{fig:HerbertSus}(a). It should generally be emphasized that there is a strong competition between $\vec{q}=0$ order and a magnetically disordered phase in this regime such that possible ordering signatures are weak and hard to identify within the PFFRG. Tracking the appearance of an instability feature during the RG flow, we find that the phase boundary between the $\vec{q}=0$ and the non-magnetic phase is approximately given by the line between $(D/J_1,J_2/J_1)=(0.04,0.08)$ and $(0.12,0)$ which goes almost through the values predicted by DFT calculations and ESR measurements for \textit{herbertsmithite}. On the paramagnetic side of the transition, the dominant spin correlations are found to be either of $\vec{q}=0$, $\sqrt{3}\times\sqrt{3}$, or \textit{incommensurate} type (i.e., at a position between $\vec{q}=0$ and $\sqrt{3}\times\sqrt{3}$ wave vectors in $\vec{k}$ space). On the other hand, the magnetic phase is completely dominated by $\vec{q}=0$ order. As an example, we show in Fig.~\ref{fig:HerbertSus} (b) the $\Lambda$ flow and the susceptibility profile for $(D/J_1,J_2/J_1)=(0.1,0.02)$. Interestingly, the latter exactly shows the type of fluctuations measured in recent neutron-scattering experiments, i.e., a large response at the Brillouin-zone boundary and small maxima at the $\vec{q}=0$ wave vectors. However, the RG flow also shows small signatures of an instability for these parameters which would possibly correspond to weak magnetic order, in contradiction with experiments. While it is difficult to draw any definite conclusion from these features, we note that such anomalies typically become more pronounced for larger system sizes and a better frequency resolution. We therefore propose the following two scenarios for \textit{herbertsmithite}: (i) The DM interaction might be smaller than the predicted value, i.e., $D/J_1\lesssim0.08$. Assuming that the Heisenberg interactions are approximately given by $J_2/J_1\approx0.02$, this would stabilize a non-magnetic phase according to our PFFRG data. Nevertheless, in these parameter regimes, PFFRG suggests that the dominant spin correlations are of $\sqrt{3}\times\sqrt{3}$ type rather than $\vec{q}=0$ which requires an additional coupling mechanism shifting the peaks. (ii) If $(D/J_1,J_2/J_1)\approx(0.1,0.02)$ describes the couplings of {\it herbertsmithite}, we find the qualitatively correct momentum profile of the spin correlations. Possible weak $\vec{q}=0$ order at these parameters could be destroyed by further frustrating interactions. Indeed, DFT simulations predict various types of ferromagnetic and antiferromagnetic interlayer couplings up to $0.035 J_1$~\cite{Jeschke2013} which could easily enhance the in-plane frustration effects. Furthermore, magnetic disorder due to copper ions on zinc sites could also be a source of quantum fluctuations in the system.

\section{Summary and conclusion}
\label{sec:Conclusion}
In this work, we have generalized the existing PFFRG approach to treat spin models with finite DM interactions. After discussing the central methodological adjustments due to off-diagonal exchanges, we tested the method for nearest-neighbor out-of-plane DM and antiferromagnetic nearest-neighbor Heisenberg interactions on the kagome lattice. We find that, at $D \geq (0.12\pm0.02) J_1$, the DM coupling destabilizes the non-magnetic phase and induces $\vec{q}=0$ order, in good agreement with exact diagonalization~\cite{Cepas2008,Seman2015}. In Sec.\ref{sec:NextNearestNeighbor}, we have further analyzed the interplay of DM interactions with first and second-neighbor Heisenberg couplings. The phase diagram of the $J_1$-$J_2$-$D$ model (see Fig.\ \ref{fig:PhaseDiagrams}) shows that, upon increasing $D$, all non-collinearly ordered phases ($\vec{q}=0$, $\sqrt{3}\times\sqrt{3}$, and cuboc orders) are enlarged while the non-magnetic phases shrink. For strong enough DM couplings ($D\gtrsim0.4J$), the non-magnetic phase around $(J_1,J_2)=(1,0)$ is found to vanish completely. Parameter regimes that have been reported to describe the mineral {\it herbertsmithite} are found to lie in close proximity to a quantum critical point between a non-magnetic phase and a $\vec{q}=0$ ordered phase. At least in parts of this parameter region, we qualitatively reproduce the low-energy neutron-scattering data from Ref.~\onlinecite{Han2016}. Despite this, the $J_1$-$J_2$-$D$ model possibly misses additional sources of frustration that might be necessary to destroy weak residual magnetic order. We argue that interlayer exchange couplings could provide such additional frustration effects.

In total, this study shows that the PFFRG approach can be successfully applied to models with finite DM couplings. Since such interactions are a consequence of lattice geometries and therefore represent a relevant perturbation in a large class of quantum magnets, we expect plenty of possibilities for future applications. For example, the next step could be to apply this technique in three spatial dimensions where it has recently been shown\cite{Iqbal2016} that the PFFRG leads to a better resolution of magnetic phase diagrams as compared to two dimensional systems.

\acknowledgments
We gratefully acknowledge discussions with Piet Wibertus Brouwer, Elina Locane, Maria Laura Baez, J\"org Behrmann, Maximilian Trescher, Christian Fr\"a\ss dorf, Ronny Thomale, and Yasir Iqbal. J.R. is supported by the Freie Universit\"at Berlin within the Excellence Initiative of the German Research Foundation.

\end{document}